# Cusp-singularity-enhanced Coriolis effect for ultrasensitive chip-scale gyroscopes

Sen Zhang[1], Dingbang Xiao[1], Fei Wang[2]*, Ran Huang[3], Lei Yu[4], Ning Zhou[4], Kaixuan He[4], Xuezhong Wu[1], Franco Nori[3]*, Hui Jing[5,6]* & Xin Zhou[1,2]*

[1]College of Intelligence Science and Technology, NUDT, Changsha, 410073, China.

[2]School of Microelectronics, Southern University of Science and Technology (SUSTech), Shenzhen, 518055, China.

[3]Center for Quantum Computing (RQC), RIKEN, Wakoshi, Saitama, 351-0198, Japan.

[4]East China Institute of Photo-Electronic IC, Bengbu, 233042, China.

[5]Institute for Quantum Science and Technology, College of Science, NUDT, Changsha, 410073, China.

[6]Key Laboratory of Low-Dimensional Quantum Structures and Quantum Control of Ministry of Education, Hunan Normal University, Changsha, 410081, China.

*Corresponding author. e-mail: wangf@sustech.edu.cn; fnori@riken.jp; jinghui73@foxmail.com; zhoux@sustech.edu.cn.

**Summary:** Gyroscopes, as fundamental inertial sensors, are crucial for rotation measurements in the consumer electronics, automotive, and aerospace industries, with the most widely used kind relying on the Coriolis effect[1-6]. The chip-scale Coriolis vibratory gyroscopes (CVGs) show reduced size, weight, and cost[1,2], but remain far lower performance than traditional macroscale CVGs[3-6], as the weak intrinsic Coriolis factor sets a fundamental limit on scaling the sensitivity against the inherently louder Brownian noise in microchips compared to the macroscale ones. Here, to overcome this physical limit, for the first time, we propose and experimentally demonstrate the use of third-order singularities lying within cusp catastrophes in the phase-tracked oscillations of an on-chip CVG to facilitate a cubic-root scaling of the Coriolis-effect-induced frequency modulation. Employing this effect, we achieve a three-order-of-magnitude enhancement in the Coriolis factor, yielding a 253-fold improvement in signal-to-noise ratio and a 297-fold increase in precision. Moreover, the cusp singularity enables a previously unattainable ultrasensitive phase-modulated sublinear measurement, achieving a world-record signal-to-noise ratio performance for silicon-chip gyroscopes. These findings not only provide revolutionary advancements in gyroscope technologies, by filling the gap in observing and controlling the singularity-enhanced Coriolis effect, but also shed new light on other ultrasensitive sensing applications.

**Main Text:**

Gyroscopes, essential sensors for measuring rotations in free space without the need for external references, play a key role in navigation and stabilization of all kinds of platforms. The most widely used type of gyroscopes operates on the same principle that governs the flight control of certain biological organisms[7,8]—the Coriolis effect, which refers to the perceived deflection of a moving object in a rotating reference frame. Known as Coriolis vibratory gyroscopes (CVGs), they sense angular rotation through Coriolis-force-induced interactions between mechanical (phononic) vibratory modes. Traditional CVGs, such as the hemispherical resonator gyroscopes (HRGs), offer high performance and reliability for applications such as oil drilling, maritime navigation, and spacecraft pointing[3-6]. However, their high cost limits their widespread use.

Lately, chip-scale CVGs have been developed for much broader uses, including movement monitoring and stabilization control of consumer electronics, automobiles, and more[1,2], due to their reduced size, weight, and cost compared to traditional CVGs. Despite these advantages, chip-scale CVGs still lag behind HRGs in performance, making them suitable only for medium- or low-end applications. Enhancing chip-scale CVG performance to match the level of the current HRGs, while preserving their miniaturization and affordability, is highly desirable to enable revolutionary technologies such as GPS-denied personal navigation, advanced robotics, and microsatellites. Yet, this goal remains elusive due to significant challenges in microfabrication errors and, more fundamentally, the Brownian noise that increases as sensor dimensions shrink, which in turn degrades the signal-to-noise ratio (SNR).

At the heart of this challenge is the limited efficiency of the Coriolis interaction, quantified by the Coriolis factor $\kappa_0$ that measures the proportion of modal mass contributing to the Coriolis effect. This factor is intrinsically determined by vibratory-mode geometry and always constrained to $\kappa_0 \leq 1$ (Supplementary Note 1). Consequently, for small rotation rates $\Omega$, the strength of the Coriolis coupling, $2\kappa_0\Omega$, is weak. The resulting physical modulations (e.g. amplitude or frequency changes) are easily blurred by the inherently stronger Brownian noise of microscale resonators in chip-scale CVGs relative to macroscale HRGs. Whether and how the Coriolis effect itself can be enhanced beyond this sensitivity limit imposed by the Coriolis factor, $\kappa_0 \leq 1$, remains an unresolved outstanding question. Resolving this barrier could enable HRG-level performance in chip-scale CVGs and unlock transformative applications.

Regarding this challenge, recent advances in singularity physics[9-13] provide new possibilities for breaking the limit of classical sensing theory, where sensing responses are proportional to a perturbation $\epsilon$. Instead, $N$th-order singularities can provide a sublinear response[14-19], $\propto \epsilon^{1/N}$, outperforming classical sensors under small perturbations ($|\epsilon| < 1$). For example, up to twentyfold sensitivity enhancements have been reported in the operation of optical gyroscopes near exceptional-point singularities[20,21], surpassing the intrinsic limits of the Sagnac effect[22]. Some singularities were also demonstrated to boost SNR in sensing[11,23-27]. However, given the inherent complexity of the Coriolis coupling in CVGs, exploiting singularity enhancement for CVGs remains an unexplored, elusive question.

Here, we address the fundamental sensitivity limit in CVG by proposing and demonstrating the enhancement of the Coriolis effect within a chip-scale CVG, achieved through the introduction and harnessing of third-order singularities residing within cusp catastrophes in a frequency-modulated (FM) operation. We incorporate additional stiffness coupling to create such cusp singularities in oscillation-frequency space facilitated by a phase-tracking control. Near the cusp singularities, we obtain a cubic-root responsivity of the singularity-enhanced Coriolis effect, which achieves a three-order magnitude increase in the Coriolis factor, a 253-fold improvement in the SNR, and a 297-fold enhancement in the precision compared to the standard FM operation, surpassing the performance enhancement factors generated from the known exceptional-point mechanisms. Moreover, the cusp singularity enables a previously unattainable ultrasensitive phase-modulated rotational measurement, achieving a bias instability of 0.035°/h and an astonishing angle random walk (ARW) of 0.00036°/√h, roughly an order-of-magnitude better than the leading silicon-chip gyroscopes, rivalling advanced HRGs of larger size and cost.

## Concept

The theoretical enhancement of the Coriolis effect by singularities starts with the typical CVG model: a proof mass supported by two orthogonal sets of springs (Fig. 1a). The oscillations of the proof mass along the red or blue springs represent standing-wave (SW) modes 1 and 2, with natural frequencies $\omega_{1,2}$ and equal dissipation rate $\gamma$. An out-of-plane rotation at angular velocity $\Omega$ induces Coriolis forces that couple the two modes, giving the CVG Hamiltonian

$$\mathbf{H} = \begin{bmatrix} \omega_1 & \mathrm{i}\kappa_0\Omega \\ -\mathrm{i}\kappa_0\Omega & \omega_2 \end{bmatrix}, \quad (1)$$

which underpins all kinds of CVG operations. The innovation in our protocol involves adding an intermodal stiffness coupling[28] $g$, implemented as the green spring in Fig. 1a (Supplementary Note 2), which, cooperating with dissipation, enables cusp singularities.

Initially, we examine a standard CVG without stiffness coupling ($g = 0$). Assuming that the modes are degenerate ($\omega_1 = \omega_2 = \omega$), the Coriolis interaction causes a shift of $\pm\kappa_0\Omega$ in the eigenfrequencies, which is harnessed to measure the angular velocity via a quadrature frequency-modulated (QFM) operation[29,30]. In this operation, a sinusoidal reference drive, $F_1$, and its $+\pi/2$ phase-shifted counterpart, $F_2$, are applied to modes 1 and 2, respectively. This induces steady-state displacements $q_{1,2} = |q_{1,2}|\cos(\omega_d t + \theta_{1,2})$, with $\omega_d$ the drive frequency and $\theta_{1,2}$ the phase lags relative to $F_1$. For $g = 0$, the quadrature drive establishes a relative mode phase $\vartheta \equiv \theta_2 - \theta_1 = \pi/2$, generating a clockwise (CW) travelling-wave (TW) mode, depicted by the circular proof-mass orbit in Fig. 1a. By tracking the $\theta_1 = -\pi/2$ contour via a phase-locked loop (PLL), we can lock into the resonance of the CW mode, as shown in Fig. 1b. The resulting oscillation frequency, $\omega + \kappa_0\Omega$, changes linearly with respect to the angular velocity, with a scale factor determined by the intrinsic Coriolis factor $\kappa_0$. This QFM operation at $g = 0$ serves as the study baseline, representing the standard Coriolis effect.

Interestingly, activating $g$ drastically changes the $\theta_1$ landscape (Fig. 1c). At zero rotation ($\Omega = 0$), the oscillation frequency tracked to the $-\pi/2$ contour of $\theta_1$, referred to as phase-tracked (PhT) frequency $\omega_T$, exhibits two "pitchfork" bifurcations at the critical coupling levels $g = (\pm\sqrt{2} - 1)\gamma$, where $\omega_T$ becomes highly nonlinear relative to $\Omega$.

The coupled dynamics is fully described by the complex mode susceptibilities $\chi_{1,2}$ Tracking the mode-1 phase $\theta_1 \equiv \mathrm{Arg}(\chi_1)$ to $-\pi/2$ gives the PhT frequency through the condition $\mathrm{Re}(\chi_1) = 0$, which leads to the cubic equation (Supplementary Note 3),

$$\delta\omega^3 + \kappa_0\Omega\delta\omega^2 + \left(\frac{\gamma^2}{4} - \frac{g\gamma}{2} - \frac{g^2}{4} - \kappa_0^2\Omega^2\right)\delta\omega - \kappa_0\Omega\left(\frac{\gamma^2}{4} + \frac{g^2}{4} + \kappa_0^2\Omega^2\right) = 0, \tag{2}$$

where $\delta\omega \equiv \omega_T - \omega$ is the frequency modulation. Plotting $\omega_T$ over the $(g, \Omega)$ space under the constraint $\mathrm{Im}(\chi_1) < 0$ yields a partially folded surface exhibiting two cusp catastrophes[13,31-37], as shown in Fig. 1d. The portion with $\mathrm{Im}(\chi_1) > 0$ (the middle sheet for $g > 0$) is omitted as it corresponds to the $\theta_1 = \pi/2$ contour and is unattainable for the $\theta_1 = -\pi/2$ PhT control. The boundary defined by $\mathrm{Im}(\chi_1) = 0$ signifies a $\theta_1$-phase singularity[38] ($PS_1$) at $g = \gamma$.

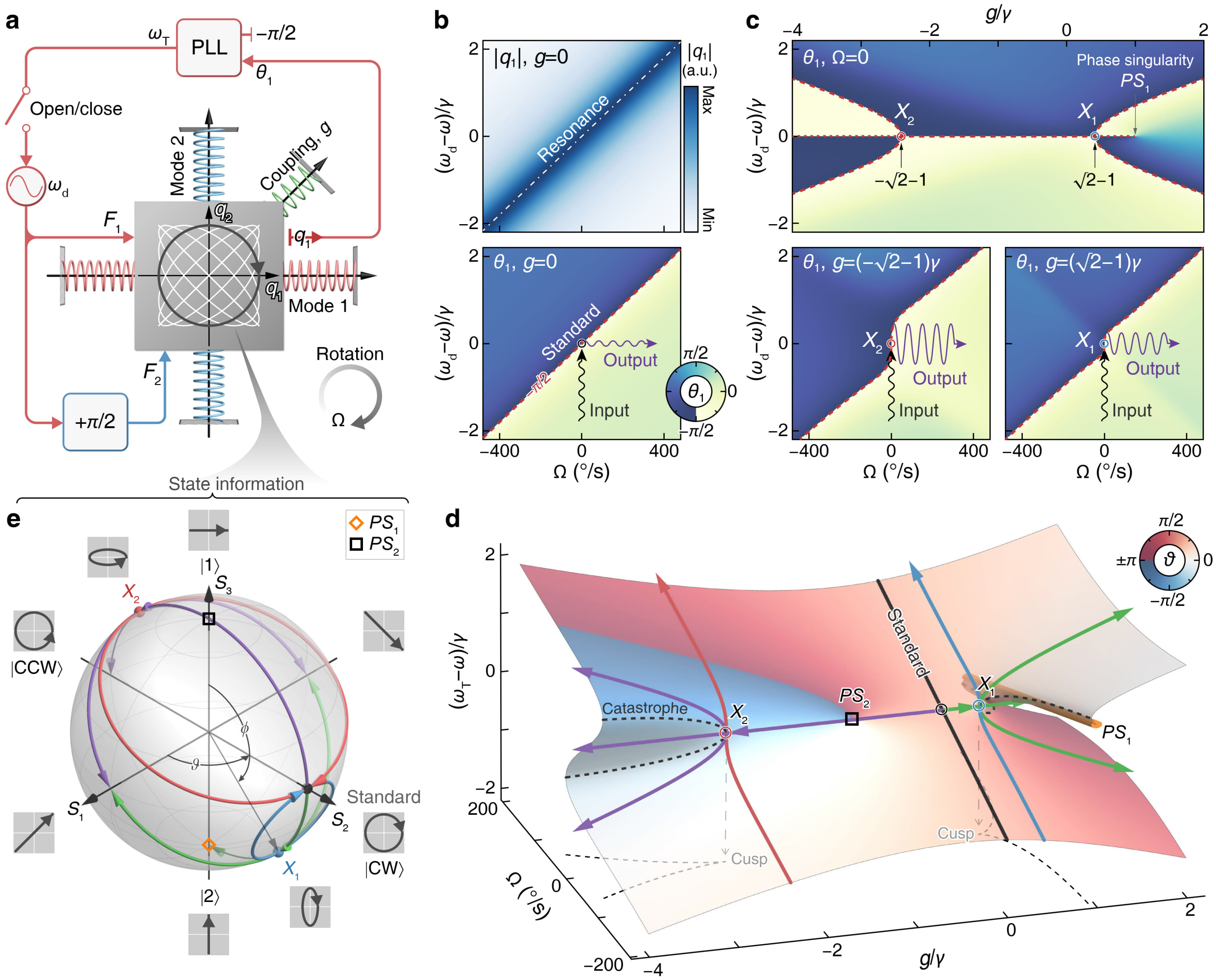

**Fig. 1. Phase-tracked (PhT) singularities and enhancement of the Coriolis effect: Concept. a**, System schematic: Degenerate modes 1 and 2, corresponding to horizontal and vertical oscillations of the proof mass, are driven by two equal-strength sinusoidal forces in quadrature, $F_{1,2}$, producing a circular mass orbit. A phase-locked loop (PLL) introduced into mode 1 adjusts the oscillation frequency to track $\theta_1$, the phase lag of mode 1 relative to $F_1$, to $-\pi/2$. Coriolis interaction induced by an out-of-plane rotation $\vec{\Omega}$ causes a shift in the oscillation frequency. To create cusp singularities, an additional coupling, $g$, is incorporated by applying a stiffness perturbation aligned between modes 1 and 2. **b**, For $g = 0$, open-loop responses of mode-1 amplitude $|q_1|$ (upper panel) and phase $\theta_1$ (lower panel) as functions of the drive frequency $\omega_d$ and angular velocity $\Omega$. The resonant frequency (dotted-dashed line) coincides with the $\theta_1 = -\pi/2$ PhT frequency $\omega_T$ (red dashed line), exhibiting a linear response to angular rotation ($\kappa_0\Omega$). **c**, Phase responses $\theta_1(\omega_d)$ as a function of coupling $g$ for zero angular velocity ($\Omega = 0$, upper panel), and as functions of the angular velocity $\Omega$ under critical coupling conditions $g = (\pm\sqrt{2} - 1)\gamma$ (lower panels), where the PhT frequency $\omega_T$ exhibits nonlinear responses to $\Omega$. **d**, PhT frequency $\omega_T$ as a function of angular velocity $\Omega$ and coupling strength $g$, showing two cusp catastrophes. The merging points of the catastrophes correspond to cusp singularities $X_{1,2}$, which are associated with two cusps when mapped onto the $g$-$\Omega$ plane. Color gradient on the surface depicts the relative phase $\vartheta \equiv \theta_2 - \theta_1$. The orange rod represents the $\theta_1$ phase singularity $PS_1$. **e**, Poincaré sphere depicting the PhT state

information under various control parameters $g$ and Ω. The colored trajectories correspond to the contours of the matching color in (**d**).

The middle (upper or lower) sheets of the folded surface are unstable (stable) (stability analysis see Supplementary Note 4). The stability boundaries constitute the catastrophes, as shown by the black dashed curves in Fig. 1d (Supplementary Note 5). Catastrophes tangentially merge at two nexuses, the signature of order-3 singularities[36]. These are referred to as cusp singularities, denoted by $X_1$ and $X_2$, as their mappings onto the $g$-Ω plane form two cusps located at $(g, \Omega) = [(\sqrt{2}-1)\gamma, 0]$ and $[(-\sqrt{2}-1)\gamma, 0]$, respectively.

For $g \neq 0$, the proof-mass orbit is no longer circular, but an elliptical hybrid mode, described by the state vector $|\psi\rangle = \cos\frac{\phi}{2}|1\rangle + e^{i\vartheta}\sin\frac{\phi}{2}|2\rangle$, where $\phi = 2\arctan(|q_2|/|q_1|)$. Projecting $|\psi\rangle$ onto a Poincaré sphere by defining spherical coordinates $S_1 = \sin\phi\cos\vartheta$ (ellipticity), $S_2 = \sin\phi\sin\vartheta$ (chirality), and $S_3 = \cos\phi$ (orientation) enables visualizing the mode hybridization (Fig. 1e and Supplementary Note 6). The QFM operation (i.e., $g = 0$) maps to the (0,1,0) point on the sphere. By examining the state evolutions following the colored contours on the $\omega_T$ surface in Fig. 1d, which are illustrated by the same-colored trajectories on the sphere, we locate the cusp singularities $X_1$ and $X_2$ at $(0,1/\sqrt{2},-1/\sqrt{2})$ and $(0,-1/\sqrt{2},1/\sqrt{2})$, respectively. The PhT states $|\psi\rangle$ can also be interpreted as hybridizations of two TW modes (Supplementary Note 7).

The cusp singularities $X_{1,2}$, possessing codimension two, are fully regulated by $g$ and Ω. Crucially, the twisted $\omega_T$ geometry near $X_{1,2}$ exhibits sharp modulations under slight changes of $g$ and Ω. An analysis shows that $X_{1,2}$ enable cubic (square) root modulations of $\omega_T$ in response to variations in Ω ($g$) (Supplementary Note 8), making $X_{1,2}$ highly suitable for the ultra-sensitive detection of Ω. In contrast to the standard Coriolis effect (lower panel of Fig. 1b), $X_{1,2}$ can generate disproportionately large changes in $\omega_T$ (outputs) for small Ω inputs (lower panels of Fig. 1c), highlighting the singular Coriolis effect.

## Experimental realization of the singularities

To verify the singular Coriolis effect and its potential for performance enhancement, we first test the experimental realization of PhT singularities by implementing the protocol in Fig. 1a to a chip-scale CVG, whose core is an electromechanical silicon disk resonator (see Methods and Extended Data Fig. 1). The resonator supports a pair of degenerate six-node SW modes 1 and 2 with a natural frequency $\omega_0 \approx 2\pi \times 40.4$ kHz and dissipation rate $\gamma \approx 2\pi \times 0.36$ Hz. These modes are electrostatically actuated, transduced, and tuned. The resonator is mounted on a temperature-controlled angular rate table to precisely manage the out-of-plane rotation. A quadrature drive applied to modes 1 and 2 excites a CW whispering gallery TW mode. The Stiffness coupling is introduced and controlled by a direct-current tuning voltage $V_c$ applied to the off-axis capacitive electrodes (see Methods).

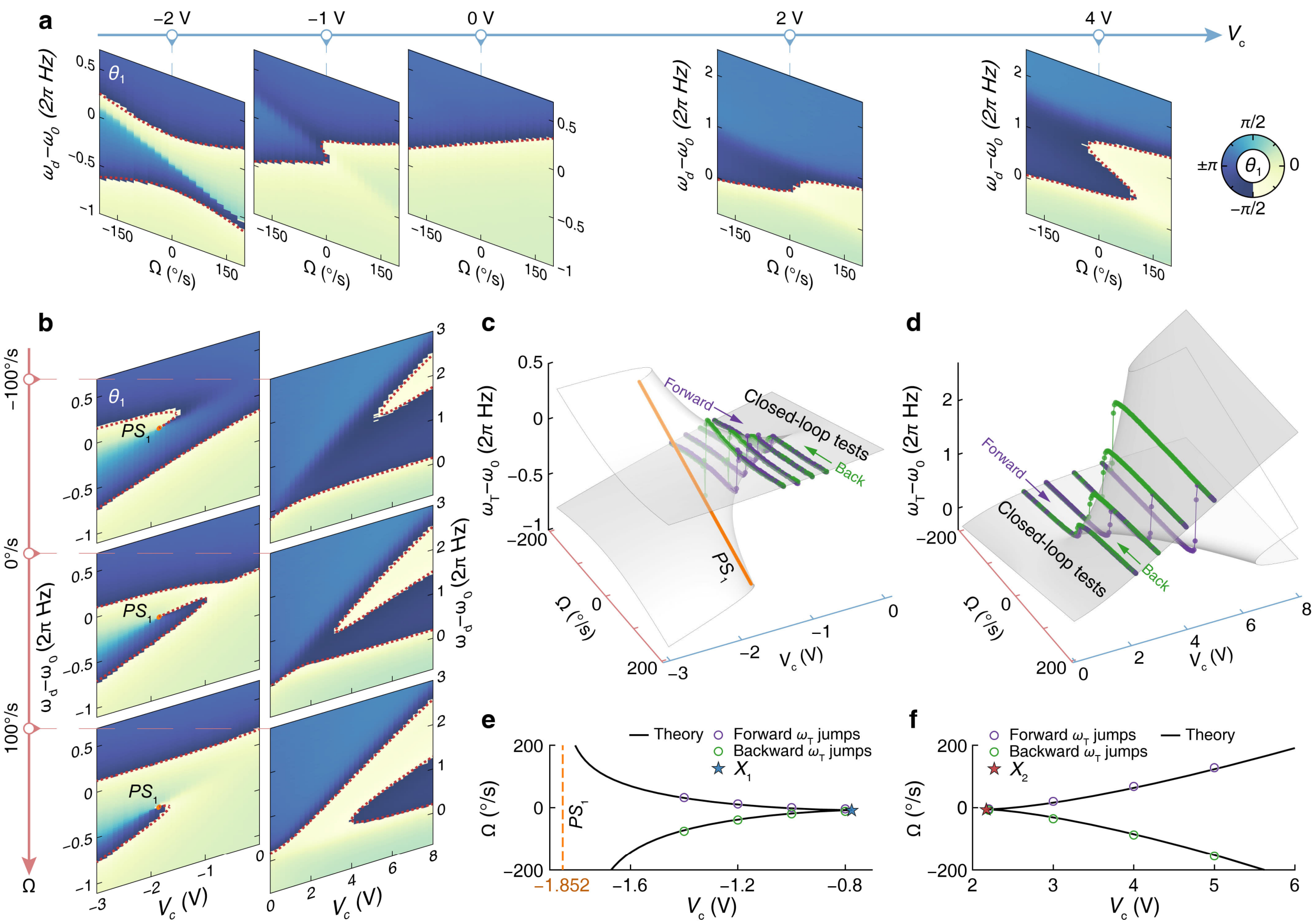

**Fig. 2. Experimental realization of PhT singularities based on a QFM operation in a chip-scale CVG. a**, Open-loop frequency responses of the mode-1 phase $\theta_1(\omega_d)$ as a function of angular velocity Ω at constant off-axis tuning voltages: $V_c \in \{-2,-1,0\}$ V and {2,4} V. In the positive (negative) $V_c$ experiments, two (one) off-axis electrodes are employed. The color edges highlight the −π/2 contours of $\theta_1$. The dotted curves are the fitting result based on the theoretical model. **b**, Open-loop frequency response of $\theta_1(\omega_d)$ as a function of the off-axis tuning voltage $V_c$ at fixed angular velocities $\Omega \in \{-100,0,100\}$°/s for negative (left panels) and positive $V_c$ (right panels). **c**,**d**, Surfaces are the theoretically computed PhT frequencies $\omega_T$ as functions of the angular velocity Ω and coupling voltage $V_c$ for (**c**) $V_c \leq 0$ and (**d**) $V_c \geq 0$. The thick orange curve marks the $\theta_1$ phase singularity $PS_1$. The purple (green) points show PLL-enabled closed-loop measurements of $\omega_T$ during forward (backward) angular velocity sweeps from −200 to 200°/s (200 to −200°/s) with 4°/s increments at fixed coupling voltages of (**c**) $V_c \in \{-1.4,-1.2,-1.0,-0.8\}$ V or (**d**)

$\{2.2,3.0,4.0,5.0\}$ V. **e**,**f**, Catastrophic frequency-jumping points (circles) of the closed-loop experiments projected onto the $V_c$-$\Omega$ plane for (**e**) negative and (**f**) positive $V_c$. The solid black curves are theoretical projections of the catastrophes showing cusp-embedded parabolas. The orange dashed line is the projection of the phase singularity $PS_1$.

To examine the $\omega_T$ surface, we characterize the $-\pi/2$ contours using the open-loop frequency responses of $\theta_1$, sweeping the drive frequency $\omega_d$ under varying $g$ and $\Omega$ conditions. We probe the contours at constant values of $V_c \in \{-2,-1,0,2,4\}$ V, and $\Omega \in \{-100,0,100\}$°/s, as shown in Fig. 2a,b. The experimental $-\pi/2$ contours of $\theta_1$ (color edges) match the theoretical model (dotted lines, Supplementary Note 3). Using the fitted parameters from Fig. 2a,b, we restore the PhT frequency $\omega_T$ as a function of $\Omega$ and $V_c$, verifying the existence of the cusp catastrophes (Fig. 2c,d).

Furthermore, we conduct closed-loop experiments with a PLL that tracks the $\theta_1 = -\pi/2$ condition, locking the drive frequency to stable branches of $\omega_T$. We perform bidirectional $\Omega$ sweeps in the range ±200°/s with 4°/s increments, under fixed $V_c \in \{-1.4,-1.2,-1.0,-0.8\}$ V or $\{2.2,3.0,4.0,5.0\}$ V. As shown in Fig. 2c and d, the closed-loop $\omega_T$ data adhere to the stable region of the open-loop-extracted $\omega_T$ surface, verifying the feasibility of the closed-loop readout of $\omega_T$.

Projecting each cusp catastrophe onto the $V_c$-$\Omega$ plane produces two parabolic branches merging at a cusp (Supplementary Note 5), as shown by the black curves in Fig. 2e and f, which are confirmed by the hysteretic transition points of the closed-loop $\omega_T$ data (circles). In the $V_c \leq 0$ case, the catastrophes do not cross the $PS_1$ at $V_c \approx -1.852$ V. Through fine-grained $\Omega$ sweeps near the hysteresis thresholds, we locate the two cusp points at $X_1$: ($V_c \approx -0.776$ V, $\Omega \approx -9.1$°/s) and $X_2$: ($V_c \approx 2.170$ V, $\Omega \approx -6.5$°/s). The slight asymmetry of $X_{1,2}$ with respect to $\Omega = 0$°/s arises from the misalignment error of the off-axis tuning electrodes, causing an unwanted natural frequency mismatch. Readjusting this mismatch repositions the cusp singularities to symmetric locations at $X_1$: ($V_c \approx -0.777$ V, $\Omega \approx 0$°/s) and $X_2$: ($V_c \approx 2.170$ V, $\Omega \approx 0$°/s) (see Methods).

## Cusp-singularity-enhanced Coriolis effect

Now, we demonstrate the enhancement of the Coriolis effect by operating the device at the previously realized PhT cusp singularities $X_{1,2}$. We measure the PhT-frequency modulations $\delta\omega_{X1,2}$ as functions of angular velocity $\Omega$ within a small range ±0.15°/s with a 0.01°/s step. Here, each output is compared against a reference recorded at a nearby, constant angular velocity to mitigate errors from the resonant frequency drift (see Methods and Extended Data Fig. 2a). The differentially measured $\delta\omega_{X1,2}$ (points in Fig. 3a) agree with the prediction (curves, Supplementary Note 8) showing a sublinear response to Ω, confirming the onset of the singular Coriolis effect. In contrast, the standard Coriolis output $\delta\omega_0$ obtained via a ±1,000°/s QFM measurements, shows a linear sensitivity of $\kappa_0 \approx 0.588$. On a logarithmic scale, the singularity-enhanced outputs have slopes of 1/3 (left side of Fig. 3b), indicating a cubic-root scaling $\delta\omega_{X1,2} \propto \Omega^{1/3}$. Whereas the standard Coriolis output maintains a linear dependence (slope of 1).

This scaling change boosts sensitivity, which is characterized by the effective Coriolis factor $\kappa \equiv \delta\omega/\Omega$, as shown in Fig. 3c. Maximal $\kappa$ of 594 and 325 are observed near $X_2$ and $X_1$, respectively, surpassing the longstanding intrinsic limit of $\kappa_0 \leq 1$. Compared with the intrinsic $\kappa_0 \approx 0.588$, a maximal sensitivity amplification factor of 1,010 (553) has been achieved for $X_2$ ($X_1$). Even at faster rotations (|Ω| >> 1°/s, right side of Fig. 3b), the singularity outputs maintain an advantage over the standard Coriolis output, despite the enhancement factors gradually diminishing with increasing |Ω| (Fig. 3c). This reconciles the trade-off between high sensitivity at small inputs and a wide dynamic range.

Then, zero-Ω bias FM outputs at $X_{1,2}$, and in the baseline (QFM, $g = 0$) configuration were recorded over a 10-minute duration to calculate the Allan deviations $\sigma_{\delta\omega}$ (Fig. 3d). The bias frequency drifts at $X_2$ ($X_1$) exceed that of the QFM baseline by 1.5 (1.3) times. This degradation, attributed to the singularity's amplification of input errors, is expected to be surpassed by the sensitivity enhancements.

Figure 3e shows the effective input deviations $\sigma_\Omega$, corresponding to the bias frequency deviations $\sigma_{\delta\omega}$, calculated using the theoretical $\omega_T$-Ω relation (Supplementary Note 10 and Extended Data Fig. 3a). Two key performance metrics are considered: the ARW, reflecting the SNR in the short term, and the bias instability, evaluating precision in the longer term. The $X_2$ ($X_1$) configuration yields a 253(37)-fold decrease in ARW and a 297(53)-fold reduction in bias instability compared with the standard QFM setup, resulting in an ARW of 0.064°/√h (0.44°/√h) and a bias instability of 2.7°/h (15.1°/h).

The experimentally demonstrated cusp-singularity-enhanced Coriolis effect in FM operation provides giant improvements in SNR and precision. This is because the dominant error source is resonant frequency fluctuations[30], which affect the FM output but are not amplified by the singularities. The influence of these errors is substantially suppressed when the FM signal is converted to the angular velocity using the singularity-enhanced sensitivities.

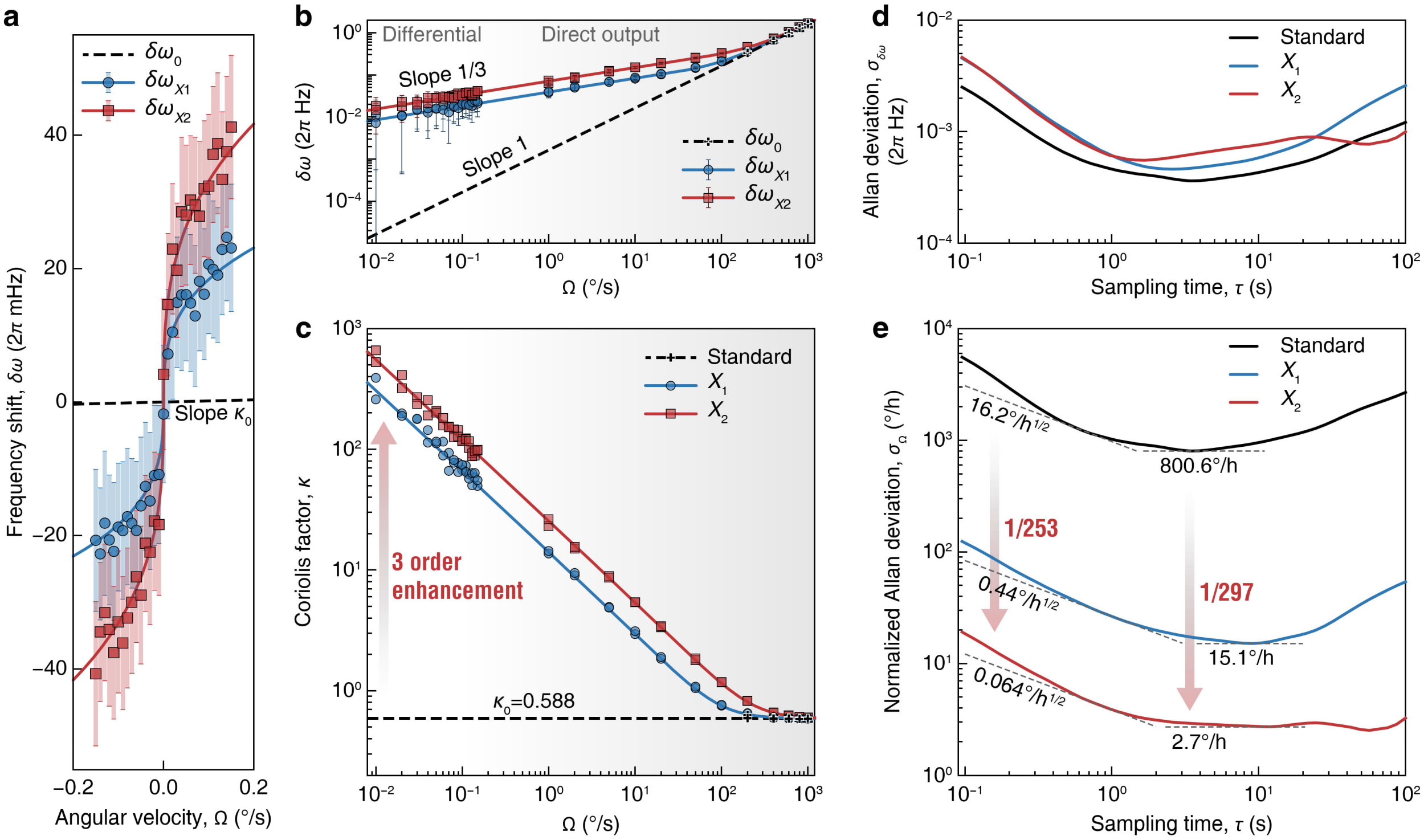

**Fig. 3. Cusp-singularity-enhanced frequency modulation. a**, Experimentally observed shifts in the PhT frequencies (red and blue points) as functions of small rotation perturbations within ±0.15°/s, stepping by 0.01°/s, when operating near the singularities $X_{1,2}$. These data are obtained using differential measurements to mitigate resonant frequency fluctuations. The red and blue curves are theoretical $\omega_T$ shifts for $g = (-\sqrt{2} - 1)\gamma$ and $(\sqrt{2} - 1)\gamma$, respectively. The black dashed line represents the fitted standard QFM Coriolis output. The error bars represent the standard deviations. **b**, Logarithmic plots of the observed frequency shifts as functions of angular velocity $\Omega$ for both slow and fast rotation measurements. Data are shown for operations initiated at singularities $X_1$ and $X_2$ (red and blue points) and standard QFM (black points). The fast-rotation ($|\Omega| > 1$°/s,) data are measured without using a differential configuration. The red, blue, and black curves are theoretical $\omega_T$ for $g = (-\sqrt{2} - 1)\gamma$, $(\sqrt{2} - 1)\gamma$, and 0, respectively. The singularity-mediated outputs exhibit slopes of 1/3, while the standard Coriolis output shows a slope of 1. **c**, Experimental (points) and theoretical (curves) Coriolis factor $\kappa \equiv \delta\omega/\Omega$ describing the sensitivity of the frequency output for singularity-mediated operations and standard operation. For minimal inputs of ±0.01°/s, the mean values of $\kappa$ are 594 (for $X_2$) and 325 (for $X_1$), representing 1,010- and 553-fold enhancements, respectively, compared with the intrinsic value $\kappa_0 \approx 0.588$. **d**,**e**, Allan deviations of (**d**) the zero-$\Omega$ bias frequency signals and (**e**) the corresponding effective bias rotations for singularity-enhanced FM and standard QFM operation. The ARW and bias instability are evaluated by the lower $\tau^{-1/2}$ and $\tau^0$ limits, respectively.

### Singularity-enabled phase-modulated measurements

Furthermore, we find that when the PhT system operates near $X_{1,2}$, the relative phase $\vartheta$ can be a superior metric for rotation readout than the PhT frequency, enabling a phase-modulated (PM) gyroscope that achieves strategic-grade SNR on silicon chips.

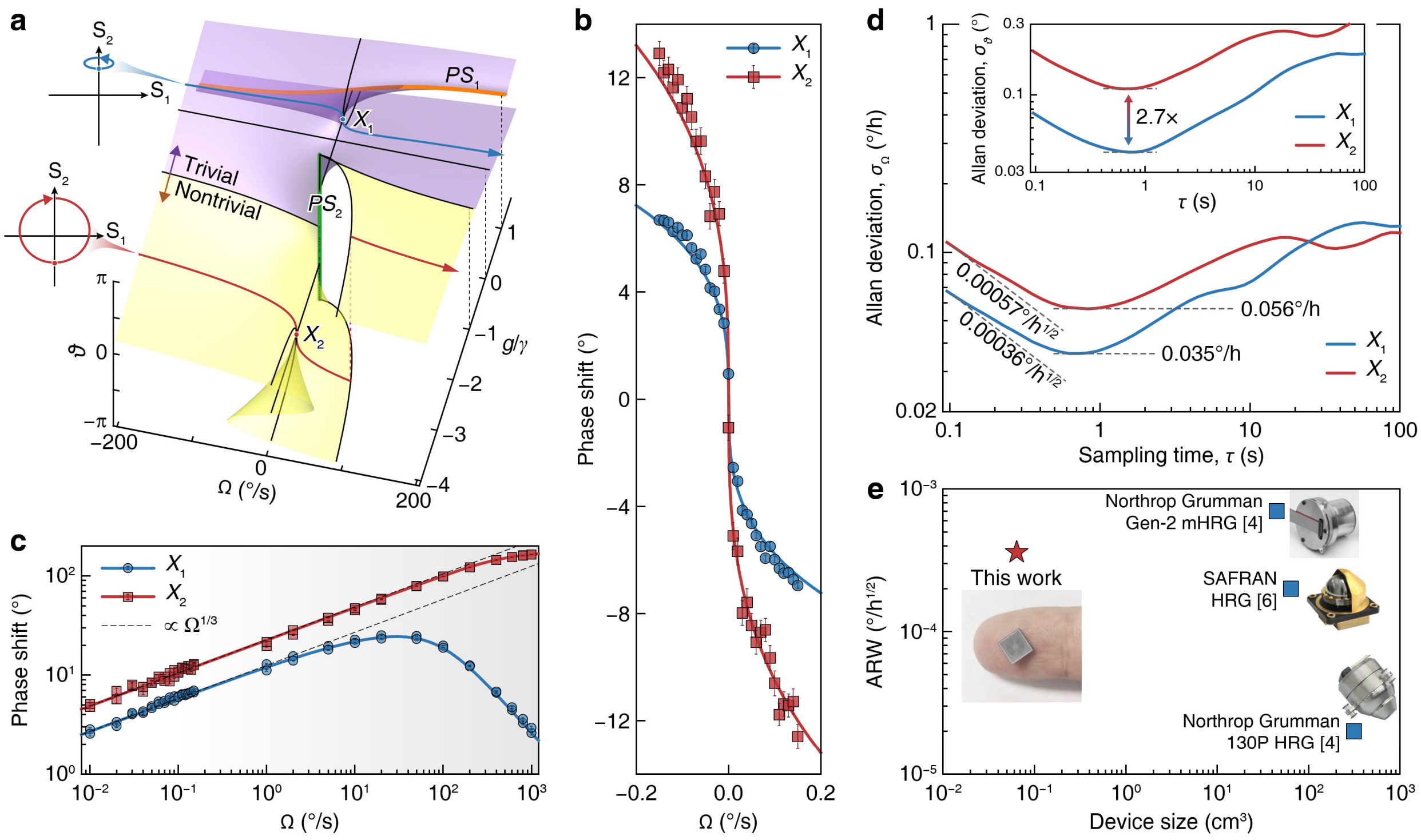

**Fig. 4. PhT singularity-mediated phase modulation for ultrasensitive rotation measurements.** **a**, Theoretically computed relative phase $\vartheta$ of the SW modes as a function of angular velocity Ω and coupling strength $g$, showing two cusp catastrophes. The cusp singularities, $X_1$ and $X_2$, are separated by a phase singularity $PS_2$ and lie within trivial and nontrivial topological phases, respectively. The state vectors projected onto the $S_1$-$S_2$ plane as Ω changes from negative to positive infinity across $X_1$ or $X_2$, are shown by the insets, depicting distinct topologies. **b**, Observed phase shift of $\vartheta$ (points) as a function of small rotation perturbations when operating near $X_{1,2}$. The phase output data are measured without using the differential configuration. The error bars represent the standard deviations. The red and blue curves are theoretical $\vartheta$ shifts for $g = (-\sqrt{2}-1)\gamma$ and $(\sqrt{2}-1)\gamma$, respectively. **c**, Logarithmic plot of the observed (points) and calculated (solid curves) phase shift of $\vartheta$ as a function of angular velocity Ω for both slow and fast rotation measurements, obtained when the system is initialized at $X_{1,2}$. **d**, Allan deviations of the effective bias inputs for the singularity-mediated zero-Ω phase outputs. Inset: Allan deviations of the original bias phase outputs. **e**, Comparison of ARW and size with some typical high-end HRGs[4,6].

The relative phase $\vartheta$ of the PhT system as a function of angular velocity $\Omega$ and coupling strength $g$ is theoretically illustrated in Fig. 4a (Supplementary Note 9). Remarkably, $\vartheta$ exhibits cusp singularities at $X_1$: $[g = (\sqrt{2} - 1)\gamma, \Omega = 0]$ and $X_2$: $[g = (-\sqrt{2} - 1)\gamma, \Omega = 0]$, coincident with those in the $\omega_T$ domain (see Fig. 1d, where the color gradient in the $\omega_T$ surface represents $\vartheta$). In standard QFM operation ($g = 0$), $\vartheta$ remains constant; whereas near the cusps it changes sharply, enabling highly sensitive PM readout.

To demonstrate the $\vartheta$-based sensing, we extract the $\vartheta$ data from the $\omega_T$ measurements in Fig. 3a. The resulting $\vartheta$-versus-$\Omega$ curves over ±0.15°/s are shown in Fig. 4b. No differential process is required because $\vartheta$ is intrinsically stable (see Extended Data Fig. 2b). On a logarithmic scale (Fig. 4c), the $\Omega$-induced phase modulations at both $X_1$ and $X_2$ exhibit slopes of 1/3, indicating a cubic-root sensitivity. As with the FM outputs, $X_2$ provides a greater sensitivity than does $X_1$.

Moreover, an intriguing topological feature emerges in the $\vartheta$ space: $X_1$ and $X_2$ are separated by a phase singularity $PS_2$. Crossing $PS_2$ induces a topological phase transition, with $X_2$ ($X_1$) lying in a nontrivial (trivial) phase. As $\Omega$ is swept from $-\infty$ to $+\infty$ through $X_2$ ($X_1$), $\vartheta$ exhibits a net winding of $-2\pi$ (0), as shown in the insets of Fig. 4a. These trajectories correspond to the red and blue paths in Fig. 1e projected onto the $S_1$-$S_2$ plane. These topological signatures are confirmed by wide-range $\Omega$ sweeps (up to ±1,000°/s) in Fig. 4c and Extended Data Fig. 2d, where the phase values are unwrapped to avoid discontinuities. This topological distinction suggests that $X_2$ supports a wider sensing range than $X_1$.

The Allan deviations of the bias $\vartheta$ signals (recorded simultaneously with FM bias signals) are shown in the inset of Fig. 4d. The bias $\vartheta$ stability at $X_1$ is about 2.7 times better than that at $X_2$, owing to the larger $|q_2|$ amplitude at $X_1$ (Extended Data Fig. 4) that reduces the measurement noise in $\theta_2$ and, consequently, in $\vartheta$. We calculate the normalized Allan deviations $\sigma_\Omega$ of the PM operation (Fig. 4d) by converting the bias $\vartheta$ signals into equivalent $\Omega$ inputs (Supplementary Note 10 and Extended Data Fig. 3b). The resulting bias instabilities reach 0.035°/h at $X_1$ and 0.056°/h at $X_2$, approaching inertial grade. Even more impressive is the SNR performance: the strategic-grade ARWs of 0.00036°/√h at $X_1$ and 0.00057°/√h at $X_2$ are comparable to those of the bulky, expensive HRGs for high-end applications[4,6] (Fig. 4e).

This PM operation delivers over two orders of magnitude improvement in both stability and SNR relative to FM operation. First, unlike $\omega_T$, $\vartheta$ is inherently unaffected by resonant-frequency fluctuations (Supplementary Note 9), thereby enabling higher measurement stability. Second, because frequency noise is integrated into phase noise, the spectral density acquires a $1/\omega_0^2$ suppression factor (Supplementary Note 11), yielding much lower noise in PM readout compared with FM detection.

Crucially, the Brownian-motion–induced phase (frequency) noise enters the PM (FM) outputs additively, without singularity amplification (Supplementary Note 11). By contrast, the singular Coriolis responses follow boosted cubic-root responsivities, enabling SNRs that exceed the fundamental limit of the conventional CVG theory. As a result, the singularity-mediated PM readout surpasses the resonator's conventional theoretical performance limits, for which the bias instability and ARW are predicted to be >0.32°/h and >0.018°/√h, respectively (Supplementary Note 12). These results establish singularity-mediated PM as a fundamentally superior transduction mechanism.

**Discussion and conclusions**

For the first time, we have demonstrated how to generate ultrasensitive responses with sublinear scaling of the Coriolis effect by operating near singularities that reside within cusp catastrophes, advancing the previous understanding that the CVG output is always proportional to the rotation input. Through this singular Coriolis effect, we have broken the physical limit of the CVG sensitivity imposed by the intrinsic Coriolis factor. We have shown that this discovery can lead to giant improvements in sensitivity, precision, and SNR of a CVG. Our findings open up fundamentally new avenues to regulating CVGs and other systems involving the Coriolis effect.

Moreover, leveraging the PM output of the enhanced Coriolis effect, we have showcased a chip-scale CVG achieving HRG-comparable, strategic grade ARW, outperforming the current cutting-edge silicon-chip gyroscopes[39-44] by almost an order of magnitude (see Extended Data Table 1 for detailed comparison). Our findings challenge the traditional view that miniaturized gyroscopes always suffer from reduced SNR, addressing the ongoing debate over whether such chip-scale gyroscopes can rival their larger traditional counterparts. Miniaturizing high-performance gyroscopes without losing precision will revolutionize the market, enabling widespread access to advanced navigation and stabilization technologies in affordable, compact devices.

Finally, the giant enhancements in sensitivity, SNR, and precision achieved through the cusp catastrophe-based singularity enabled by PhT control—among the highest in recent singularity-enhanced sensing experiments[11,16-26] (see Extended Data Table 2 for detailed comparison)—along with the ultrasensitive singularity-enabled PM operation demonstrated in this study, could lead to advancements in any field requiring extreme sensitivity. The PhT cusp-singularity-enhanced paradigm can be adapted to all kinds of sensing applications, such as environmental monitoring[45], healthcare sensing[46], seismology[47], gravity measurement[48], and even gravitational-wave detection[49], potentially revolutionizing the development of more sensitive, compact, and cost-effective measurement systems.

At present, the demonstrated cusp-singularity–enhanced measurements employ a single-channel, non-self-calibrated configuration. Future implementations that incorporate differential architectures[29,30,50] could further improve bias stability.

**References**

1. R. Antonello, R. Oboe, "MEMS gyroscopes for consumers and industrial applications", in *Microsensors*, I. Minin, Ed. (IntechOpen, Rijeka, 2011) Chap. 12.
2. C. Acar, A. Shkel, *MEMS Vibratory Gyroscopes: Structural Approaches to Improve Robustness*, MEMS Reference Shelf (Springer New York, NY, 2009).
3. E. C. Litty, L. L. Gresham, P. A. Toole, D. A. Beisecker, Hemispherical resonator gyro: an IRU for Cassini. In *Proc. SPIE-The International Society for Optical Engineering* **2803**, 299 (SPIE, 1996).
4. D. M. Rozelle, A. D. Meyer, A. A. Trusov, D. K. Sakaida, Milli-HRG inertial sensor assembly–a reality. In *Proc. 2015 IEEE International Symposium on Inertial Sensors and Systems (INERTIAL)* 1–4 (2015).
5. J. Campanile, Enhanced scaleable SIRU. In *Proc. 2016 IEEE/ION Position, Location and Navigation Symposium (PLANS)* **86**, 905–909 (IEEE, 2016).
6. F. Delhaye, HRG by SAFRAN: The game-changing technology. In *Proc. 2018 IEEE International Symposium on Inertial Sensors and Systems (INERTIAL)* 1–4 (IEEE, 2018).
7. W. P. Chan, F. Prete, M. H. Dickinson, Visual input to the efferent control system of a fly's "gyroscope". *Science* **280**, 289 (1998).
8. S. P. Sane, A. Dieudonné, M. A. Willis, T. L. Daniel, Antennal mechanosensors mediate flight control in moths. *Science* **315**, 863 (2007).
9. M.-A. Miri, A. Alù, Exceptional points in optics and photonics. *Science* **363**, eaar7709 (2019).
10. S. K. Ozdemir, S. Rotter, F. Nori, L. Yang, Parity-time symmetry and exceptional points in photonics. *Nature Materials* **18**, 783 (2019).

11. R. Kononchuk, J. Feinberg, J. Knee, T. Kottos, Enhanced avionic sensing based on Wigner's cusp anomalies. *Science Advances* **7**, eabg8118 (2021).

12. K. Peters, S. Rodriguez, Exceptional precision of a nonlinear optical sensor at a square-root singularity. *Physical Review Letters* **129**, 013901 (2022).

13. X. Zhou, X. Ren, D. Xiao, J. Zhang, R. Huang, Z. Li, X. Sun, X. Wu, C.-W. Qiu, F. Nori, H. Jing, Higher-order singularities in phase-tracked electromechanical oscillators. *Nature Communications* **14**, 7944 (2023).

14. J. Wiersig, Enhancing the sensitivity of frequency and energy splitting detection by using exceptional points: Application to microcavity sensors for single-particle detection. *Physical Review Letters* **112**, 203901 (2014).

15. Z.-P. Liu, J. Zhang, S. K. Ozdemir, B. Peng, H. Jing, X.-Y. Lu, C.-W. Li, L. Yang, F. Nori, Y.-X. Liu, Metrology with PT-symmetric cavities: Enhanced sensitivity near the PT-phase transition. *Physical Review Letters* **117**, 110802 (2016).

16. W. Chen, S. K. Ozdemir, G. Zhao, J. Wiersig, L. Yang, Exceptional points enhance sensing in an optical microcavity. *Nature* **548**, 192 (2017).

17. H. Hodaei, A. U. Hassan, S. Wittek, H. Garcia-Gracia, R. El-Ganainy, D. N. Christodoulides, M. Khajavikhan, Enhanced sensitivity at higher-order exceptional points. *Nature* **548**, 187 (2017).

18. J. Wiersig, Prospects and fundamental limits in exceptional point-based sensing. *Nature Communications* **11**, 2454 (2020).

19. J. Wiersig, Review of exceptional point-based sensors. *Photonics Research* **8**, 1457–1467 (2020).

20. M. P. Hokmabadi, A. Schumer, D. N. Christodoulides, M. Khajavikhan, Non-Hermitian ring laser gyroscopes with enhanced Sagnac sensitivity. *Nature* **576**, 70 (2019).

21. Y.-H. Lai, Y.-K. Lu, M.-G. Suh, Z. Yuan, K. Vahala, Observation of the exceptional-point-enhanced Sagnac effect. *Nature* **576**, 65 (2019).

22. W. W. Chow, J. Gea-Banacloche, L. M. Pedrotti, V. E. Sanders, W. Schleich, M. O. Scully, The ring laser gyro. *Reviews of Modern Physics* **57**, 61 (1985).

23. R. Kononchuk, J. Cai, F. Ellis, R. Thevamaran, T. Kottos, Exceptional-point-based accelerometers with enhanced signal-to-noise ratio. *Nature* **607**, 697 (2022).

24. A. Suntharalingam, L. Fernandez-Alcazar, R. Kononchuk, T. Kottos, Noise resilient exceptional-point voltmeters enabled by oscillation quenching phenomena. *Nature Communications* **14**, 5515 (2023).

25. J. Xu, Y. Mao, Z. Li, Y. Zuo, J. Zhang, B. Yang, W. Xu, N. Liu, Z. J. Deng, W. Chen, K. Xia, C.-W. Qiu, Z. Zhu, H. Jing, K. Liu, Single-cavity loss-enabled nanometrology. *Nature Nanotechnology* **19**, 1472 (2024).

26. Y.-P. Ruan, J.-S. Tang, Z. Li, H. Wu, W. Zhou, L. Xiao, J. Chen, S.-J. Ge, W. Hu, H. Zhang, C.-W. Qiu, W. Liu, H. Jing, Y.-Q. Lu, K. Xia, Observation of loss-enhanced magneto-optical effect. *Nature Photonics* **19**, 109 (2025).

27. A. Suntharalingam, L. Fernández-Alcázar, P. F. Wagner-Boián, M. Reisner, U. Kuhl, T. Kottos, Symmetry-violation-driven hysteresis loops as measurands for noise-resilient sensors. *Physical Review Applied* **23**, 064043 (2025).

28. S.-B. Shim, M. Imboden, P. Mohanty, Synchronized oscillation in coupled nanomechanical oscillators. *Science* **316**, 95-99 (2007).

29. M. H. Kline, Y.-C. Yeh, B. Eminoglu, H. Najar, M. Daneman, D. A. Horsley, B. E. Boser, Quadrature FM gyroscope. In *Proc. 2013 IEEE 26th International Conference on Micro Electro Mechanical Systems (MEMS)* 604–608 (IEEE, 2013).

30. M. Kline, Frequency modulated gyroscopes. (University of California, Berkeley, CA, 2015).

31. R. Thom, *Structural stability and morphogenesis* (Benjamin, 1975).

32. V. I. Arnold, *Catastrophe Theory* (Springer-Verlag, 1984).

33. M. G. Castellano, F. Chiarello, R. Leoni, F. Mattioli, G. Torrioli, P. Carelli, M. Cirillo, C. Cosmelli, A. de Waard, G. Frossati, N. Grønbech-Jensen, S. Poletto, Catastrophe observation in a Josephson-junction system. *Physical Review Letters* **98**, 177002 (2007).

34. S. Rosenblum, O. Bechler, I. Shomroni, R. Kaner, T. Arusi-Parpar, O. Raz, B. Dayan, Demonstration of fold and cusp catastrophes in an atomic cloud reflected from an optical barrier in the presence of gravity. *Physical Review Letters* **112**, 120403 (2014).

35. J. Mumford, E. Turner, D. W. L. Sprung, D. H. J. O'Dell, Quantum spin dynamics in Fock space following quenches: Caustics and vortices. *Physical Review Letters* **122**, 170402 (2019).

36. W. Tang, X. Jiang, K. Ding, Y.-X. Xiao, Z.-Q. Zhang, C. T. Chan, G. Ma, Exceptional nexus with a hybrid topological invariant. *Science* **370**, 1077 (2020).

37. J. del Pino, J. J. Slim, E. Verhagen, Non-Hermitian chiral phononics through optomechanically induced squeezing. *Nature* **606**, 82 (2022).

38. J. Ni, C. Huang, L.-M. Zhou, M. Gu, Q. Song, Y. Kivshar, C.-W. Qiu, Multidimensional phase singularities in nanophotonics. *Science* **374**, eabj0039 (2021).

39. A. D. Challoner, H. G. Howard, J. Y. Liu, Boeing disc resonator gyroscope. In *Proc. IEEE/ION Position, Location and Navigation Symposium (PLANS)* 504–514 (IEEE/ION, 2014).

40. S. Askari, M. Asadian, K. Kakavand, A. Shkel, Near-navigation grade quad mass gyroscope with Q-factor limited by thermo-elastic damping. In *Proc. Solid-State Sensors, Actuators and Microsystems Workshop (Hilton Head)* 254–257 (2016).

41. D. Endean, K. Christ, P. Duffy, E. Freeman, M. Glenn, M. Gnerlich, B. Johnson, J. Weinmann, Near-navigation grade tuning fork MEMS gyroscope. In *Proc. 2019 IEEE International Symposium on Inertial Sensors and Systems (INERTIAL)* 1–4 (IEEE, 2019) .

42. S. Koenig, S. Rombach, W. Gutmann, A. Jaeckle, C. Weber, M. Ruf, D. Grolle, J. Rende, Towards a navigation grade SiMEMS gyroscope. In *Proc. 2019 DGON Inertial Sensors and Systems (ISS)* 1–18 (IEEE, 2019).

43. M. Gadola, A. Buffoli, M. Sansa, A. Berthelot, P. Robert, G. Langfelder, 1.3 $mm^2$ nav-grade NEMS-based gyroscope. *Journal of Microelectromechanical Systems* **30**, 513 (2021).

44. L. Chen, Q. Li, T. Miao, P. Wang, X. Zhang, Y. Zhang, X. Wu, D. Xiao, 0.003°/h bias instability of honeycomb disk resonator gyroscope achieved by mode reversal combined mode deflection control method. *Microsystems & Nanoengineering* **11**, 152 (2025).

45. J. S. Apte, C. Manchanda, High-resolution urban air pollution mapping. *Science* **385**, 380 (2024).

46. C. S. Wood, M. R. Thomas, J. Budd, T. P. Mashamba-Thompson, K. Herbst, D. Pillay, R. W. Peeling, A. M. Johnson, R. A. McKendry, M. M. Stevens, Taking connected mobile-health diagnostics of infectious diseases to the field. *Nature* **566**, 467 (2019).

47. P. Lognonné *et al.*, SEIS: Insight's seismic experiment for internal structure of Mars. *Space Science Reviews* **215**, 12 (2019).

48. R. P. Middlemiss, A. Samarelli, D. J. Paul, J. Hough, S. Rowan, G. D. Hammond, Measurement of the Earth tides with a MEMS gravimeter. *Nature* **531**, 614 (2016).

49. B. P. Abbott *et al.* (LIGO Scientific Collaboration and Virgo Collaboration), Observation of gravitational waves from a binary black hole merger. *Physical Review Letters* **116**, 061102 (2016).

50. B. Eminoglu, Y.-C. Yeh, I. I. Izyumin, I. Nacita, M. Wireman, A. Reinelt, B. E. Boser, Comparison of long-term stability of AM versus FM gyroscopes. In *Proc. 2016 IEEE 29th International Conference on Micro Electro Mechanical Systems (MEMS)* 954–957 (IEEE, 2016).

## Methods

### Experimental realization

The gyroscopic system outlined in Fig. 1a is realized using a 4 mm-diameter on-chip disk resonator exhibiting twelve-fold symmetry. Extended Data Fig. 1a displays a top-view photograph, highlighting the resonator in yellow. The resonator's deformable structure features ten concentric rings linked by radial spokes, with its central part bonded to a substrate. Fabricated from 100 μm-thick P-type single-crystal silicon, the resonator is housed in a vacuum environment to reduce air damping. The resonator and associated circuitry are integrated on a printed circuit board mounted on a temperature-controlled (25±0.05°C) angular rate table. This setup allows accurate out-of-plane rotations with precise angular velocities (error <0.001°/s). Electric connections are maintained during rotation using slip rings, linking the rotating parts with external equipment.

The elastic deformations in the rings and spokes of the silicon disk resonator lead to multiple eigenstates. This research utilizes a pair of nearly degenerate in-plane wineglass standing-wave (SW) modes with wave number of $n = 3$ (i.e., with $2 \times n = 6$ nodes or anti-nodes arranged circularly), designated as modes 1 and 2 in Extended Data Fig. 1b. Each standing-wave mode is a combination of two in-plane whispering-gallery traveling-wave (TW) modes sharing the wave number $n = 3$, which travel in opposite directions, denoted as clockwise (CW) and counterclockwise (CCW) in Extended Data Fig. 3c. The relationship between standing-wave and traveling-wave modes is detailed later on.

The fixed nodes and anti-nodes of standing-wave modes enable actuation, transduction, and tuning using fixed capacitive electrodes, situated with uniform 10 μm gaps from the resonator. Electrodes are color-coded by function in Extended Data Fig. 1a. Differential actuation is achieved by applying anti-phase signals to two opposite anti-nodal electrodes. Modes 1 and 2 are driven simultaneously, with mode 2's drive signal phase-shifted by $+\pi/2$ relative to mode 1. The quadrature excitation of these modes creates a CW traveling-wave mode. Anti-nodal displacements of modes 1 and 2 are detected capacitively in a differential setup using charge amplifiers.

The transduced displacement signals, $q_{1,2}$, are processed by a lock-in amplifier (Zurich Instruments MFLI) for demodulation against the reference driving signal, yielding the amplitudes $|q_{1,2}|$ and phases $\theta_{1,2}$. The relative phase $\vartheta$ between $\theta_2$ and $\theta_1$ is also recorded. Open-loop measurements use the lock-in amplifier's parametric sweeper block, sequentially adjusting the reference driving frequency $\omega_d$ while monitoring the amplitudes and phases as functions of $\omega_d$, i.e., $|q_{1,2}|(\omega_d)$, $\theta_{1,2}(\omega_d)$. In closed-loop operation, a phase-locked loop (PLL) with a PID controller keeps the phase of mode 1 at $-\pi/2$, by adjusting the oscillation frequency. This PLL-controlled frequency, achieving $\theta_1 = -\pi/2$ phase-tracking, is denoted as $\omega_T$.

Imperfections in the degeneracy of the fabricated disk resonator are corrected through a preparatory tuning process utilizing a proven mode-matching technique[51]. This procedure employs two electrode sets, depicted as pink and green in Extended Data Fig. 1a. A DC voltage $V_c$ is implemented on one (for negative $V_c$) or two (for positive $V_c$) off-axis green electrodes between the principal axes of modes 1 and 2 to facilitate stiffness coupling. This results in an electrostatically tunable off-axis spring with stiffness $\Delta_c = E_{1,2}\,(2V_0V_c - V_c^2)$, where $E_1 \approx$ 10,066 N m$^{-1}$kg$^{-1}$V$^{-2}$ or $E_2 \approx$ 21,068 N m$^{-1}$kg$^{-1}$V$^{-2}$ correspond to tuning factors for one or two electrodes[52]. Here, $V_0 = 30$ V is the static voltage applied to the resonator body. The off-axis tuning introduces a coupling with strength $g = -\Delta_c/(2\omega_0)$, with the negative sign due to a roughly −15° azimuthal angle between the tuning electrodes and the principal axis of mode 1. Fabrication flaws may cause slight misalignment of the electrodes with the central axis of modes 1 and 2, resulting in minor non-degeneracy, which is corrected by readjusting the in-axis tuning voltage $V_f$ on the pink electrodes in Extended Data Fig. 1a. The DC voltages $V_0$, $V_c$, $V_f$ are provided by precise IT2800 source measure units.

## Sensitivity measurements

In the small-range measurement, the system is carefully tuned at the balanced cusp singularities, $X_1$: ($V_c \approx -0.777$ V, $\Omega \approx 0$°/s) or $X_2$: ($V_c \approx 2.170$ V, $\Omega \approx 0$°/s), to enhance gyroscopic sensitivity. The angular velocity, $\Omega$, is varied incrementally from −0.15°/s to 0.15°/s in 0.01°/s steps, with stabilization at each point for about 120 seconds to allow the system to settle. The steady-state outputs, phase-tracked frequency $\omega_T$ and relative phase $\vartheta$, are recorded for 5 seconds, producing 525 datapoints for each angular velocity, depicted by blue (for $X_1$) or red (for $X_2$) points in Extended Data Fig. 2a and b for $\omega_T$ and $\vartheta$, respectively.

The $\omega_T$ frequency outputs in small-range measurements are significantly affected by resonant-frequency fluctuations. To mitigate these effects, a differential output configuration is used, measuring the difference between each frequency output and a reference output $\omega_T(\Omega_r)$ at a constant angular velocity $\Omega_r = 1.5°/s$ after each valid output (black points in Extended Data Fig. 2a). These differential values are adjusted by $\omega_T(\Omega_r) - \omega_T(0)$ to shift the origin to zero, with $\omega_T(0)$ being the expected frequency at $\Omega = 0°/s$. The offset frequency differences $\delta\omega_{X1,2}$ at singularities $X_1$ and $X_2$ are indicated by blue and red points in Fig. 3a, respectively. Each point represents the average of 525 samples, with error bars showing the standard deviation.

In contrast, the phase outputs $\vartheta$ for small-range measurements exhibit much greater stability, as shown in Extended Data Fig. 2b. Thus, no differential operation is applied to the $\vartheta$ outputs, notwithstanding that reference outputs $\vartheta(\Omega_r)$ have been recorded (black points in Extended Data Fig. 2b). The mean and standard deviation of 525 $\vartheta$ samples for each angular velocity input are depicted as points with error bars in Fig. 4b.

For extensive measurements, the angular velocity is sequentially set to $\Omega \in \{0,\mp1,\mp2,\mp5,\mp10,\mp20,\mp50,\mp100,\mp200,\mp400,\mp600,\mp800,\mp1000\}°/s$, following system calibration to balanced cusp singularities. The stabilized phase-tracked frequency $\omega_T$ and relative phase $\vartheta$ are maintained for 5 seconds to capture 525 data points at each angular velocity. Mean values for $\omega_T$ and $\vartheta$ across angular velocities are represented by blue ($X_1$) and red ($X_2$) points in Extended Data Fig. 2c and d, with standard deviations as error bars. The $\vartheta$ values are unwrapped to prevent discontinuities. Similarly, measurements are performed with the standard QFM at $V_c = 0$, using angular velocities $\Omega \in \{0,\mp200,\mp400,\mp600,\mp800,\mp1000\}°/s$, recording only $\omega_T$, shown by black points in Extended Data Fig. 2c and Fig. 3b.

**Data availability**

All data are available on figshare[53] (https://doi.org/10.6084/m9.figshare.29278061.v3). Source data are provided with this paper.

51. X. Ren, X. Zhou, Y. Tao, Q. Li, X. Wu, D. Xiao, Radially pleated disk resonator for gyroscopic application. *Journal of Microelectromechanical Systems* **30**, 825 (2021).
52. X. Zhou, C. Zhao, D. Xiao, *et al.* Dynamic modulation of modal coupling in microelectromechanical gyroscopic ring resonators. *Nature Communications* **10**, 4980 (2019).
53. S. Zhang *et al.* Enhancement of the Coriolis effect via cusp singularities for ultrasensitive chip-scale gyroscopes, dataset. *figshare* https://doi.org/10.6084/m9.figshare.29278061.v3 (2025).

**Acknowledgments**

X.Z. thank Prof. Ashwin Seshia from the University of Cambridge for helpful discussion. This work is partly supported by the National Key R&D Program grants 2024YFE0102400 (H.J.) and 2022YFB3204901 (X.Z.), the National Natural Science Foundation of China (NSFC) grants U21A20505 (D.X., X.Z., and X.W.), 11935006 (H.J.), 12421005 (H.J.), and 62174077 (F.W.), the Hunan Major Sci-Tech Program grant 2023ZJ1010 (H.J.), the RIKEN Special Postdoctoral Researchers (SPDR) program (R.H.), Young Elite Scientist Sponsorship Program by CAST grant YESS20200127 (X.Z.), and the Natural Science Foundation of Hunan Province for Excellent Young Scientists grant 2021JJ20049 (X.Z.). This work is primarily supported by the National Natural Science Foundation of China (NSFC) grant 52575679 (X.Z.).

**Author contributions**

X.Z. conceived the idea. F.N., H.J., and X.Z. initiated the research. X.Z. and S.Z. performed the experiments. X.Z. processed the data. X.Z., H.J., and F.N. conducted the theory. X.Z. designed the device. L.Y., N.Z., K.H., and X.Z. fabricated the device. X.Z., S.Z., F.W., D.X., and X.W. developed the test system. X.Z., H.J., and F.N. wrote the manuscript with inputs from all authors. H.J., F.N., R.H., F.W., and X.Z. revised the manuscript. H.J., F.N., and X.Z. jointly supervised the project.

**Competing interests**

The authors declare no competing interests.

**Additional information**

**Correspondence and requests for materials** should be addressed to Franco Nori, Hui Jing, or Xin Zhou.

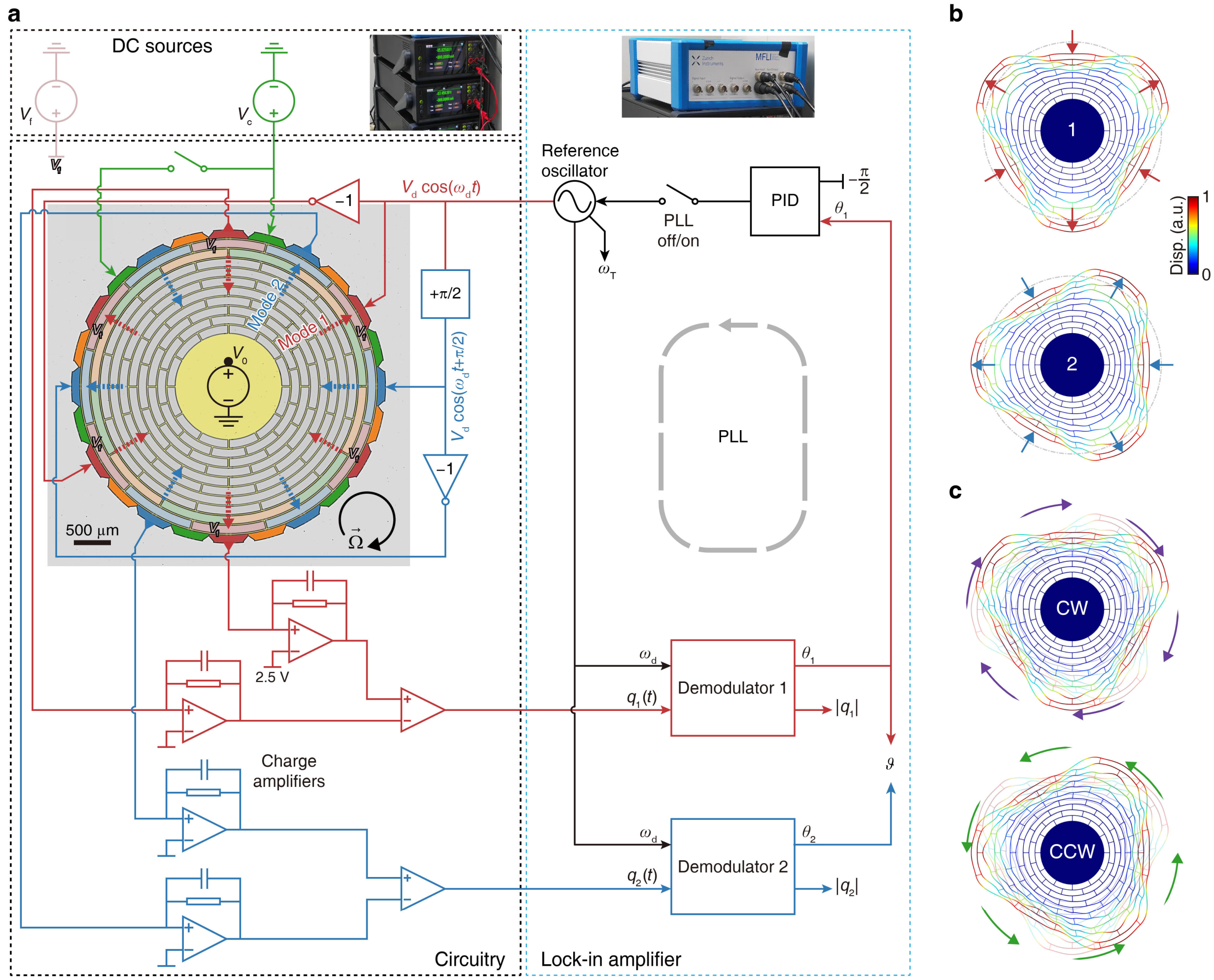

**Extended Data Fig. 1. Experimental realization.** **a**, Image and schematic of the experimental configuration. The picture of the disk-resonator device is a pseudo-coloured microscope photograph. The device, along with its signal processing circuitry, is embedded on a printed circuit board mounted on an angular rate table. A lock-in amplifier facilitates drive, detection, and PLL control. Electrostatic tuning is accomplished through the application of precise direct-current (DC) voltages. PLL, phase-locked loop. **b**, Mode shapes with exaggerated amplitudes for standing-wave modes 1 and 2. **c**, Diagram of CW and CCW travelling-wave modes.

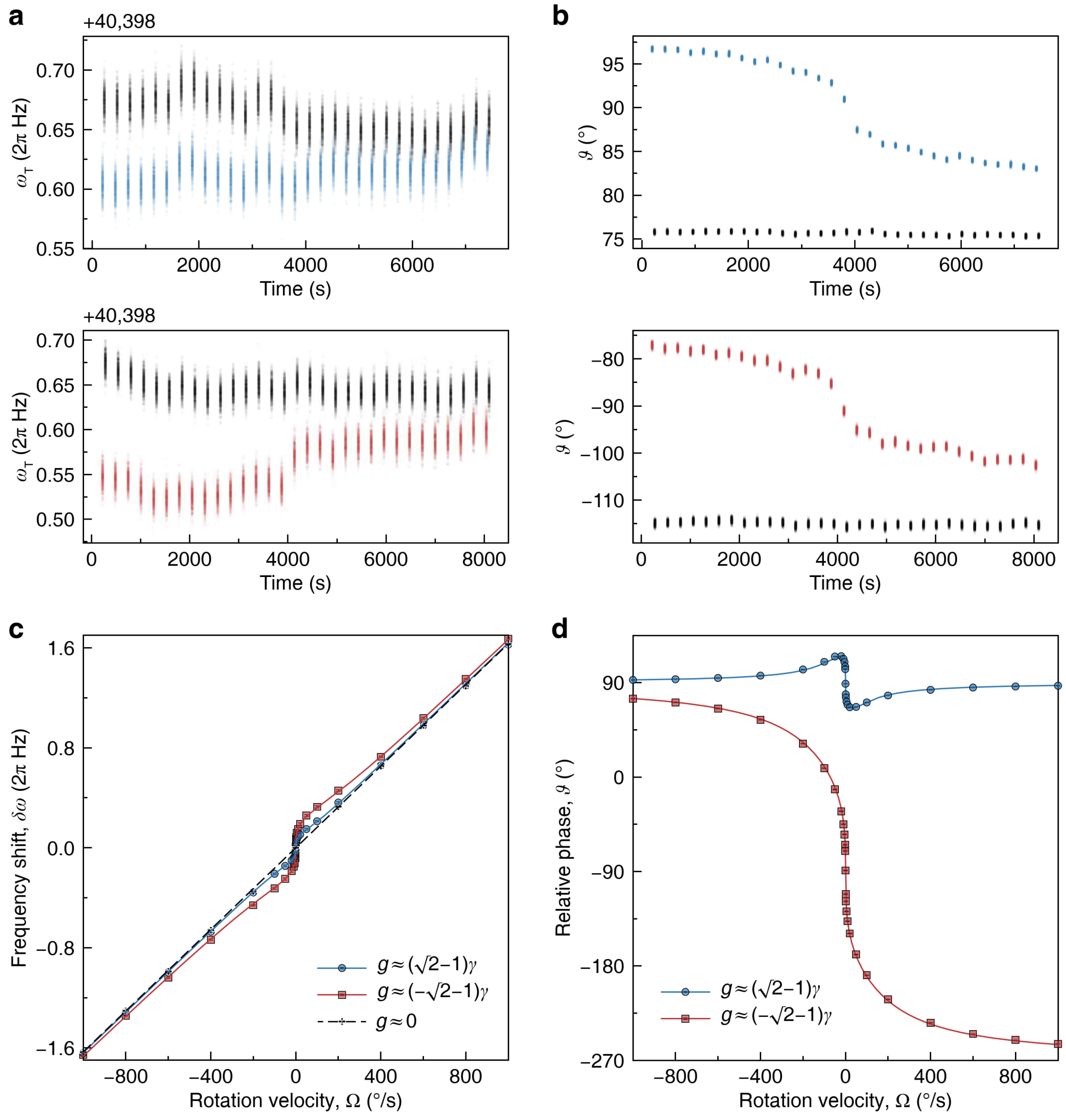

**Extended Data Fig. 2. Measurements of the phase-tracked frequency and relative phase as dependents on angular velocity. a**, The phase-tracked frequency $\omega_T$ as a function of angular velocity $\Omega \in \{-0.15,-0.14,\ldots,0,\ldots,0.14,0.15\}$°/s with the system calibrated to balanced cusp singularities $X_1$ (blue) and $X_2$ (red). Each cluster comprises 525 data points of $\omega_T$ at the same $\Omega$ input. Black markers represent $\omega_T(\Omega_r)$ measured at a constant reference angular velocity $\Omega_r$ = 1.5°/s. **b**, Relative phase $\vartheta$ corresponding to the measurements in (**a**). **c**, Shifts in the phase-tracked frequency $\omega_T$ as a function of angular velocity in large-range measurements when the system is tuned to operate at $X_1$ $[g \approx (\sqrt{2}-1)\gamma]$, $X_2$ $[g \approx (-\sqrt{2}-1)\gamma]$, or standard quadrature frequency-modulated mode ($g \approx 0$). **d**, Relative phase $\vartheta$ corresponding to the measurements in (**c**) operated at $X_{1,2}$. The error bars represent the standard deviations.

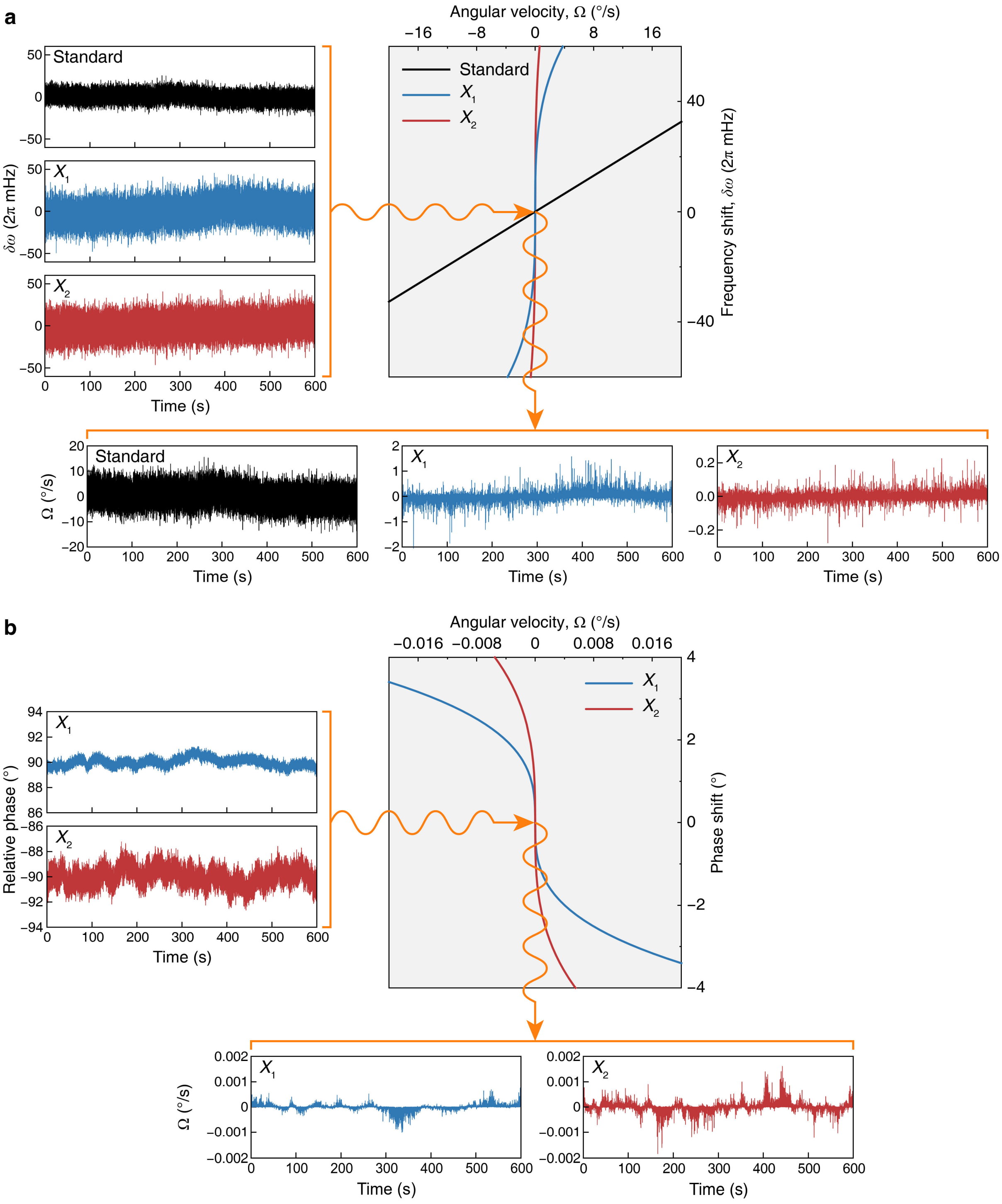

**Extended Data Fig. 3. Bias outputs and estimated effective angular-velocity inputs. a**, Zero-rotation bias signals of the phase-tracked frequency $\omega_T$ when the system is operated in standard quadrature frequency-modulated mode and at the balanced cusp singularities $X_{1,2}$. Effective angular-velocity inputs corresponding to the bias output signals are estimated using the Ω to $\omega_T$ transduction functions. **b**, Bias signals of the relative phase $\vartheta$ corresponding to the $\omega_T$-bias measurements at $X_{1,2}$ in (**a**). Effective inputs are estimated using the Ω to $\vartheta$ transduction functions. The Ω estimation processes are detailed in Supplementary Note 10.

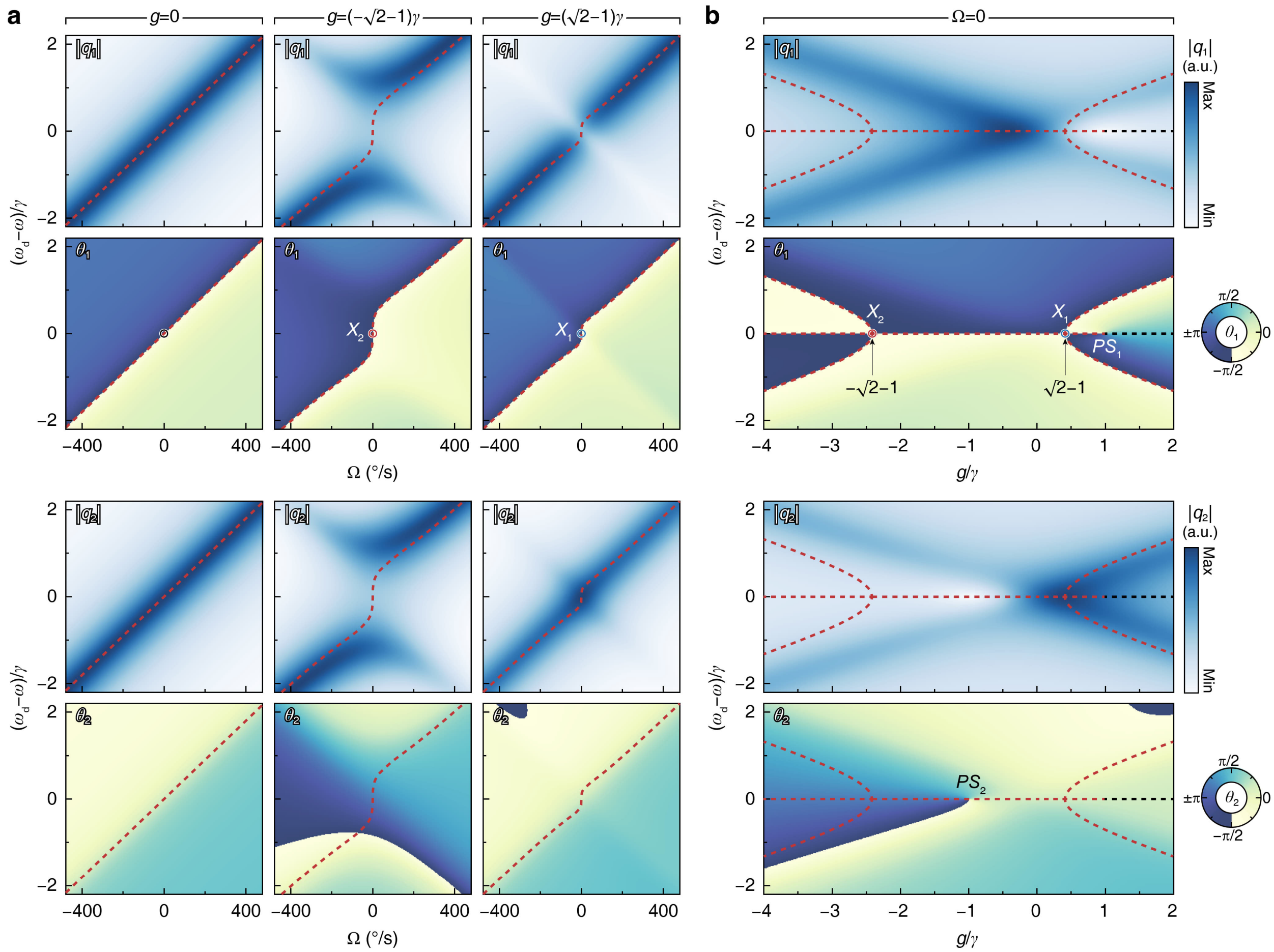

**Extended Data Fig. 4. Theoretical frequency responses of standing wave modes, $|q_{1,2}|(\omega_d)$ and $\theta_{1,2}(\omega_d)$. a**, The amplitudes $|q_{1,2}|(\omega_d)$ and phases $\theta_{1,2}(\omega_d)$ are depicted as functions of $\Omega$ under the functions $g = 0$, $(-\sqrt{2} - 1)\gamma$ and $(\sqrt{2} - 1)\gamma$. The $|q_{1,2}|(\omega_d)$ and $\theta_{1,2}(\omega_d)$ are shown as functions of $g$ at fixed $\Omega = 0$. The red (black) dashed lines denote the equiphase contours where $\theta_1 = -\pi/2$ ($\pi/2$).

**Extended Data Table 1. Comparison to state-of-the-art high-performance silicon-chip Coriolis vibratory gyroscopes.**

| Reference | Config. | Structure | Resonant Frequency, Detuning (if any) | Quality factor | Intrinsic Coriolis factor | Proof mass | Drive Ampl. | ARW (°/√h) | Bias stability (°/h) | Year |
|---|---|---|---|---|---|---|---|---|---|---|
| Boeing, JPL [39] | AM | Disk | 14 kHz | 80k | 0.8 | 1.1 mg | 1.8 μm | 0.0021 | 0.012 (Allan) | 2014 |
| University of California, Irvine [40] | AM | Quad-mass tuning fork | 1.6 kHz | 1.8M | 0.87 | 8 mg | 2.5 μm | 0.015 | 0.09 (Allan) | 2016 |
| Honeywell [41] | AM | Dual-mass tuning fork | --, >700 Hz | -- | >0.9 | -- | -- | 0.003 | 0.009 (Allan)<br>0.05 (1σ@10s) | 2019 |
| Northrop Grumman LITEF GmbH [42] | AM | Dual-mass tuning fork | -- | -- | >0.9 | -- | -- | 0.02 | 0.007 (Allan)<br>0.3 (1σ@10s) | 2019 |
| Politecnico di Milano [43] | AM | Dual-mass tuning fork | 25 kHz, 100-200 Hz | 25k-50k | >0.9 | 3.5 μg | 9 μm | 0.004 | 0.02 (Allan) | 2021 |
| Silicon Sensing CRS39A | AM | Ring | 14 kHz | -- | 0.8 | -- | -- | 0.004 | 0.03 (Allan) | 2021 |
| NUDT [44] | AM | Disk | 4.2k | 570k | 0.85 | 1.8 mg | 2 μm | 0.003 | 0.015 (Allan, standard),<br>0.003 (Allan, mode reversal&deflection)<br>0.08 (1σ@10s, mode reversal&deflection) | 2025 |
| This work | Singularity enhanced | Disk | 40.4kHz | 112k | 0.588 | 60 μg | 0.5 μm | 0.00036 | 0.035 (Allan) | 2025 |

**Extended Data Table 2. Comparison to previous experiments of singularity-enhanced sensing.**

| Reference | Singularity type (order) | Degrees of Freedom | Enhancement object | Enhancements | | | Year |
|---|---|---|---|---|---|---|---|
| | | | | Responsivity | SNR | Precision | |
| Chen et.al. [16] | Exceptional point (2) | 2 | Nanoparticle sensing | 2.5× | -- | -- | 2017 |
| Hodaei et.al. [17] | Exceptional point (3) | 3 | Joule heat sensing | 23× | -- | -- | 2017 |
| Hokmabadi et.al. [20] | Exceptional point (2) | 2 | Sagnac effect | 20× | -- | -- | 2019 |
| Lai et.al. [21] | Exceptional point (2) | 2 | Sagnac effect | 4× | Ineffective | Ineffective | 2019 |
| Kononchuk et.al. [11] | Wigner's cusp anomaly (2) | 2 | Acceleration sensing | 60× | 8.4× | 7.8× | 2021 |
| Kononchuk et.al. [23] | Exceptional point (2) | 2 | Acceleration sensing | 10× | 3× | ~3× | 2022 |
| Suntharalingam et.al. [24] | Nonlinear exceptional point (2) | 2 | Voltage sensing | 100× | >150× | >4.5× | 2023 |
| Xu et.al. [25] | Exceptional point (2) | 2 | Nanometrology | 86× | 5× | ~8× | 2024 |
| Ruan et.al. [26] | Exceptional point (2) | 2 | Magneto-optical effect | 10× | 3× | -- | 2024 |
| This work | Cusp singularity (3) | 2 | Coriolis effect | 1010× | 253× | 297× | 2025 |